\documentclass[%
 reprint,
 amsmath,amssymb,
 aps,
]{revtex4-2}

\usepackage{graphicx}
\usepackage{dcolumn}
\usepackage{bm}
\usepackage{color}

\begin{document}


\title{H-T phase diagrams of a holographic p-wave superfluid}

\author{Yu-Ni Yang}
\affiliation{Kunming University of Science and Technology, Kunming 650500, China}
\author{Chuan-Yin Xia}
\affiliation{Kunming University of Science and Technology, Kunming 650500, China}
\author{Zhang-Yu Nie}
 \email{niezy@kust.edu.cn}
 \email{Corresponding author.}
\affiliation{Kunming University of Science and Technology, Kunming 650500, China}


\author{Hua-Bi Zeng}
\affiliation{Center for Gravitation and Cosmology, College of Physical Science and Technology, Yangzhou University, Yangzhou 225009, China}


\date{\today}

\begin{abstract}
We study the competition between the p-wave and the p+ip superfluid solutions in a holographic model with applied magnetic field intensity $H$. We find that when $H$ is turned on, both the grand potential and the critical temperature of the p+ip solution are shifted, while the p-wave solution is only slightly affected. Combining the effect of $H$ and back reaction parameter b, we build $H-T$ phase diagrams with a slit region of p+ip phase. The zero (or finite) value of $H$ at the starting point of the slit region is related to second (or first) order of the p-wave phase transition at zero magnetic intensity, which should be universal in systems with degenerate critical points (spinodal points) at zero magnetic field.
\begin{description}
\item[Usage]
xxx
\item[Structure]
xxx
\end{description}
\end{abstract}

\maketitle


\section{\label{sec:introduction}Introduction}
The gauge/gravity duality~\cite{Maldacena:1997re,Gubser:1998bc,Witten:1998qj} provides a useful tool to investigate the strongly coupled field theory. Since the success of holographic modeling of superfluid phase transition~\cite{Gubser:2008px,Hartnoll:2008vx,Herzog:2008he}, this duality has been applied to various condensed matter systems~\cite{Zaanen:2015oix}.

Different from the s-wave superfluid with a complex scalar order, the superfluid with triplet pairing are more complicated. The holographic studies on p-wave superfluid~\cite{Gubser:2008wv,Cai:2013aca} also reveal interesting features, such as the anisotropic conductivity~\cite{Gubser:2008wv}, which is different to the s-wave superfluid. In the first study on holographic p-wave superfluid~\cite{Gubser:2008wv}, both the U(1) gauge field and the charged vector orders are realized by SU(2) gauge fields, where the charge and mass parameters of the vector order are fixed to 1 and 0 respectively. In the new holographic p-wave model~\cite{Cai:2013aca}, the authors using charged complex vectors, which are also known as Proca fields, to extend the holographic p-wave superfluid model to more general values of the charge and the mass parameters. Therefore a lot of study emerge to consider the more general p-wave superfluid in various setups~\cite{Cai:2013pda,Li:2013rhw,Wu:2014bba,
Rogatko:2015nta,Pan:2015lit,Lai:2016yma,Nie:2016pjt,Lu:2018tdo,Lu:2020phn,Huang:2019yov,Lv:2020ecm,Nie:2020lop,Qiao:2020hkx,Lu:2021tln}.
Further investigations also involve in  both the s-wave and p-wave orders together to study the competition and coexistence between the two different orders in holography~\cite{Nie:2013sda,Amado:2013lia,Momeni:2013bca,Nie:2014qma,Liu:2015zca,
Nie:2015zia,Wang:2016jov,Li:2017wvw,Momeni:2019kfl,Xia:2021pap,Zhang:2021vwp}.

Besides the usual anisotropic p-wave order, a new kind of p+ip order with isotropic energy momentum tensor in $x-y$ plane is also valid. The p+ip order is firstly studied holographically in the SU(2) p-wave model~\cite{Gubser:2008wv}, but the p+ip solution turns out to be unstable even in probe limit. Several years later, the p+ip solution with applied magnetic field is studied near the linear region in the new holographic p-wave model~\cite{Cai:2013aca} with charged complex vector fields in bulk~\cite{Cai:2013pda}. Until Ref.~\cite{Nie:2016pjt}, the p+ip solution with finite condensate is realized in the new holographic p-wave model~\cite{Cai:2013aca}, and is discovered to be the same stable with the p-wave solution in the probe limit. However, the p+ip solution is less stable than the p-wave solution with considering the back-reaction on metric, and further study in DRGT massive gravity also failed to stabilize the p+ip solution beyond the probe limit~\cite{Nie:2020lop}.

Although the p+ip order is isotropic in x-y plane, it breaks the chiral symmetry. Therefore it is possible to increase the stability of the p+ip solution by introducing coupling to applied magnetic field. Study on real world p-wave superfluid such as Helium-3 also indicates that the applied magnetic field indeed has nontrivial influence on the phase structure. Furthermore, with a stabilized p+ip phase, it is convenient to study interesting features such as the chiral conductivity. Therefore in this paper, we study the competition between the p-wave and p+ip orders with applied magnetic field in the new holographic p-wave model~\cite{Cai:2013aca}. We hope to get p+ip phase that is more stable than the p-wave solution, and expect interesting $H-T$ phase diagrams.

The rest of this paper is organized as follows. In Sec.~\ref{sect:setup}, we give the setup of the holographic model. In Sec.~\ref{sect:b} we review the influence of back-reaction on the p-wave and p+ip phase transitions studied in Ref.~\cite{Nie:2016pjt}. In Sec.~\ref{sect:H}, we study the influence of the magnetic field intensity $H$ on the phase transitions. In Sec.~\ref{sect:PD}, we combine the effects from back-reaction and magnetic field to build two typical $H-T$ phase diagrams with a narrow wedge region dominated by the p+ip phase, we also discuss on a possible universality for a slit region in phase diagram in this section. We give discussions and conclusions in Sec.~\ref{sect:conclusion}.

\section{The holographic setup}\label{sect:setup}
We start with the complex vector field holographic p-wave model~\cite{Cai:2013aca,Cai:2013pda,Li:2013rhw,Wu:2014bba,Pan:2015lit,Lai:2016yma,Nie:2014qma} and work in 3+1 dimensional bulk spacetime. The action of this model is
\begin{eqnarray}
S  &=&S_G+S_M,\\
S_G&=&\frac{1}{2 \kappa_g ^2}\int d^{4}x \sqrt{-g} \big(R-2\Lambda \big),\\
S_M&=&\frac{1}{q_p^2}\int d^{4}x \sqrt{-g}\Big(-\frac{1}{4}F_{\mu\nu}F^{\mu\nu} -\frac{1}{4}M_{\mu\nu}M^{\mu\nu} \nonumber\\
&&-\frac{1}{2} \rho^\dagger_{1\mu\nu}\rho_1^{\mu\nu}
-m_p^2 \rho^\dagger_{1 \mu} \rho_1^\mu \nonumber+i q_p \gamma \rho_{1 \mu} \rho^\dagger_{1 \nu} M^{\mu\nu} \nonumber\\
&&-\frac{1}{2} \rho^\dagger_{2 \mu\nu}\rho_2^{\mu\nu}
-m_p^2 \rho^\dagger_{2 \mu} \rho_2^\mu -\lambda \rho^\dagger_{1 \mu} \rho_1^\mu \rho^\dagger_{2 \nu} \rho_2^\nu\Big).
\end{eqnarray}
We write the action of the holographic model into sum of the gravity section $S_G$ and the matter section $S_M$. $F_{\mu\nu}=\nabla_\mu A_\nu-\nabla_\nu A_\mu$ is the field strength for the U(1) gauge field $A_\mu$, while $M_{\mu\nu}=\nabla_\mu B_\nu-\nabla_\nu B_\mu$ is the filed strength of magnetic field $B_{\mu}$. In 3+1 dimensional bulk spacetime There exist dyonic solution in which both the magnetic and electric fields are non-zero. However, we separate the electric and magnetic parts into two different fields to simplify the numerical work when considering back reaction of condensed vector fields with non-zero magnetic field.

$\rho_{1\mu}$ and $\rho_{2\mu}$ are two complex vector fields charged under the U(1) gauge field with charge $q$, and the superscript ``$^\dagger$" means complex conjugate. The field strength of $\rho_{1\mu}$ and $\rho_{2\mu}$ are $\rho_{j\mu\nu}=D_\mu \rho_{j\nu}-D_\nu \rho_{j\mu}, (j=1,2)$ with covariant derivative $D_\mu =\nabla_\mu -i q A_\mu$. $m_p$ is the mass for the complex vector fields and controls the conformal dimension of the p-wave orders in the dual field theory. We choose the values of $q$ and $m_p$ to be the same for the two vector orders. The only difference between them is that $\rho_{1\mu}$ is coupled to the magnetic strength $M_{\mu\nu}$, while  $\rho_{2\mu}$ is magnetically neutral. We use $\rho_{1\mu}$ to realize the p+ip order while use $\rho_{2\mu}$ to realize the p-wave order.

The last term in the action is an interaction term between the two vector fields, this term will only influence the coexistent solution and do not change the p-wave or p+ip solution with single condensate, therefore we use its coefficient $\lambda$ to control the stability of the coexistence solution and make the final phase diagram self consistent.

In AdS$^4$, when magnetic field is turned on, the vector field $\rho_{1\mu}$ coupled to the magnetic field only permits the p+ip solution with both the x and y components non-zero. In order to also include the p-wave phase in our study, we have to introduce the magnetic neutral vector $\rho_{2\mu}$ with the same value of $q$ and $m_p$. If we work in AdS$^5$, the p-wave solution can be consistently realized from the third boundary direction component of the vector field $\rho_{1\mu}$, and the holographic model can be simplified by removing the second vector $\rho_{2\mu}$. However, we can take advantage of the analytic dyonic black hole solution in AdS$^4$ to simplify the numerical calculation, where the magnetic field $B_\mu$ is fixed to $B_y=H x$, with considering the full gravitational dynamics consistently.

The equations of motion for the full system can be given by the equations for matter fields
\begin{eqnarray}
\nabla^\nu F_{\nu\mu} = i  (\rho_1^\nu\rho_{1 \nu\mu}^\dagger-\rho_1^{\nu\dagger}\rho_{1 \nu\mu}) \nonumber\\
+i  (\rho_2^\nu\rho_{2 \nu\mu}^\dagger-\rho_2^{\nu\dagger}\rho_{2 \nu\mu}),\\
\nabla^\nu M_{\nu\mu} = i q_p \gamma \nabla^\nu (\rho_{1\nu}\rho_{1 \mu}^\dagger-\rho_{1\nu} ^\dagger \rho_{1 \mu}), \\
D^\nu \rho_{1 \nu\mu}  - m_p^2 \rho_{1 \mu} +i q_p \rho_{1 \nu} {M^\nu}_\mu -\lambda\rho_{1\mu}\rho_{2 \nu}^\dagger\rho_2^\nu= 0,\\
D^\nu \rho_{2 \nu\mu}  - m_p^2 \rho_{2 \mu} -\lambda\rho_{1 \nu}^\dagger\rho_1^\nu\rho_{2 \mu}= 0,
\end{eqnarray}
and the Einstein equations
\begin{equation}
R_{\mu\nu} -\frac{1}{2}(R-2\Lambda) g_{\mu\nu} = b^2 \mathcal{T}_{\mu\nu},
\end{equation}
where $b= \kappa_g/q$ characterizes the strength of back reaction of the matter fields on the metric fields. $\mathcal{T}_{\mu\nu}$ is
the stress-energy tensor of the matter sector
\begin{eqnarray}
&&\mathcal{T}_{\mu\nu} = (-\frac{1}{4}F_{\alpha\beta}F^{\alpha\beta} -\frac{1}{4}M_{\alpha\beta}M^{\alpha\beta}  \\
&&-\frac{1}{2} \rho^\dagger_{1\alpha\beta}\rho_1^{\alpha\beta}
-m_p^2 \rho^\dagger_{1 \alpha} \rho_1^\alpha+i q_p \gamma \rho_{1 \alpha} \rho^\dagger_{1 \beta} M^{\alpha\beta} \nonumber\\
&&-\frac{1}{2} \rho^\dagger_{2 \alpha\beta}\rho_2^{\alpha\beta}
-m_p^2 \rho^\dagger_{2 \alpha} \rho_2^\alpha -\lambda \rho^\dagger_{1 \alpha} \rho_1^\alpha \rho^\dagger_{2 \beta} \rho_2^\beta) g_{\mu\nu} \nonumber\\
&& +F_{\mu\lambda} F_\nu^\lambda +M_{\mu\lambda} M_\nu^\lambda -2i q_p \gamma (\rho_{1\mu}\rho_1^{\dagger\lambda}M_{\nu\lambda}+\rho_1^\lambda\rho^\dagger_{1\mu}M_{\lambda\nu}) \nonumber\\
&& + \rho^\dagger_{1\mu\lambda}\rho_{1\nu}^\lambda+\rho^\dagger_{1\nu\lambda}\rho_{1\mu}^\lambda+m_p^2 (\rho^\dagger_{1\mu} \rho_{1\nu}+ \rho^\dagger_{1\nu} \rho_{1\mu}) \nonumber\\
&& + \rho^\dagger_{2\mu\lambda}\rho_{2\nu}^\lambda+\rho^\dagger_{2\nu\lambda}\rho_{2\mu}^\lambda+m_p^2 (\rho^\dagger_{2\mu} \rho_{2\nu}+ \rho^\dagger_{2\nu} \rho_{2\mu}) \nonumber\\
&&-\lambda (\rho^\dagger_{1\mu} \rho_{1\nu}+ \rho^\dagger_{1\nu} \rho_{1\mu})\rho^\dagger_{2 \alpha} \rho_2^\alpha
-\lambda \rho^\dagger_{1 \alpha} \rho_1^\alpha(\rho^\dagger_{2\mu} \rho_{2\nu}+ \rho^\dagger_{2\nu} \rho_{2\mu}). \nonumber
\end{eqnarray}

According to the previous analysis, we use $\rho_{1\mu}$ to realize the p+ip solution and use $\rho_{2\mu}$ to realize the p-wave solution. The ansatz for the matter fields can be taken as
\begin{eqnarray}
A_t=\phi(r),~B_y=H x,~ \nonumber\\
\rho_{1x}=\Psi_x(r),~\rho_{1y}=i \Psi_y(r),~\rho_{2x}=\Psi_p(r),
\end{eqnarray}
with all other field components set to zero. A consistent metric form~\cite{Nie:2014qma,Ammon:2009xh,Nie:2016pjt} is
\begin{eqnarray}\label{fullmetric}
&ds^2=&-N(r) \sigma (r)^2dt^2+\frac{1}{N(r)}dr^2 \\ \nonumber
&&+\frac{r^2}{L^2}(\frac{1}{f(r)^2}dx^{2}+f(r)^2 dy^2),
\end{eqnarray}
with
\begin{equation}
N(r)=\frac{r^2}{L^2}(1-\frac{4 M(r)+b^2 H^2 L^6}{2 r^3}+\frac{b^2 H^2 L^6}{2 r^4}),
\end{equation}
where $L$ is the AdS radius and related to the cosmological constant by $\Lambda=-3/L^2$.

With the above matter and metric ansatz, the equations of motion can be written as
\begin{eqnarray}
M'(r)&&=\frac{b^2 L^4 N}{2} \Big( f^2 \Psi_p'^2 +f^2 \Psi_x'^2 +\frac{\Psi_y'^2}{f^2}\Big) \nonumber \\
&&+\frac{b^2 L^4 q_p^2 \phi ^2}{2 N(r) \sigma ^2} \Big( f^2 \Psi_p^2 +f^2 \Psi_x^2 +\frac{\Psi_y^2}{f^2}\Big) \nonumber \\
&&+\frac{b^2 L^4 m_p^2}{2} \Big( f^2 \Psi_p^2 +f^2 \Psi_x^2 +\frac{\Psi_y^2}{f^2}\Big) \nonumber \\
&&-\frac{b^2L^6 q_p \gamma H}{r^2} \Psi_x \Psi_y + \frac{b^2L^6 \lambda}{2 r^2} \Psi_p^2 \Big(f^4\Psi_x^2+\Psi_y^2\Big) \nonumber \\
&&+\frac{b^2 L^2 r^2 \phi '^2}{4 \sigma ^2} +\frac{L^2 r^2 N f'^2}{2 f^2},
\end{eqnarray}
\begin{eqnarray}
\sigma '(r)&&=\frac{r \sigma f'^2}{f^2}
+\frac{b^2 L^2 \sigma}{r}\Big(f^2\Psi_p'^2 +f^2\Psi_x'^2 +\frac{\Psi_y'^2}{f^2} \Big)  \nonumber\\
&&+\frac{b^2 L^2 q_p^2 \phi^2}{r N^2 \sigma}\Big(f^2 \Psi_p^2 +f^2 \Psi_x^2+\frac{\Psi_y^2}{f^2} \Big),
\end{eqnarray}
\begin{eqnarray}
f''(r)&&=\frac{f'^2}{f}-\frac{f' N'}{N}-\frac{f'\sigma '}{\sigma}-\frac{2 f'}{r}  +\frac{2b^2L^4f^5 \lambda}{r^4 N} \Psi_p^2 \Psi_x^2    \nonumber\\
&&+\frac{b^2 L^2 f}{r^2}\Big(f^2\Psi_p'^2 +f^2\Psi_x'^2 -\frac{\Psi_y'^2}{f^2} \Big) \nonumber\\
&&-\frac{b^2 L^2 f q_p^2\phi^2}{r^2 N^2 \sigma^2} \Big(f^2 \Psi_p^2 +f^2 \Psi_x^2-\frac{\Psi_y^2}{f^2} \Big) \nonumber \\
&&+\frac{b^2 L^2 f m_p^2}{r^2 N} \Big(f^2 \Psi_p^2 +f^2 \Psi_x^2-\frac{\Psi_y^2}{f^2} \Big) ,
\end{eqnarray}
\begin{eqnarray}
\phi''(r)&&= \Big(\frac{\sigma '}{\sigma}-\frac{2 }{r} \Big)\phi ' \nonumber \\
&&+\frac{2 L^2 q_p^2}{r^2 N}\Big(f^2 \Psi_p^2 +f^2 \Psi_x^2+\frac{\Psi_y^2}{f^2} \Big)  \phi,
\end{eqnarray}
\begin{eqnarray}
\Psi_p''(r)&& = -\Big( \frac{N'}{N}+\frac{\sigma'}{\sigma}+\frac{2 f'}{f} \Big) \Psi_p'
-\Big(\frac{q_p^2 \phi^2}{N^2 \sigma^2} -\frac{m_p^2}{N}\Big) \Psi_p \nonumber \\
&&+\frac{L^2 \lambda}{r^2 N} \Big(f^2 \Psi_x^2+\frac{\Psi_y^2}{f^2} \Big) \Psi_p,
\end{eqnarray}
\begin{eqnarray}
\Psi_x''(r)&& = -\Big( \frac{N'}{N}+\frac{\sigma'}{\sigma}+\frac{2 f'}{f} \Big) \Psi_x'
-\Big(\frac{q_p^2 \phi^2}{N^2 \sigma^2} -\frac{m_p^2}{N}\Big) \Psi_x \nonumber \\
&&+\frac{L^2 \lambda}{r^2 N} f^2 \Psi_p^2 \Psi_x  - \frac{L^2 q_p \gamma H \Psi_y }{r^2 f^2 N},
\end{eqnarray}
\begin{eqnarray}
\Psi_y''(r)&& = -\Big( \frac{N'}{N}+\frac{\sigma'}{\sigma}-\frac{2 f'}{f} \Big) \Psi_y'
-\Big(\frac{q_p^2 \phi^2}{N^2 \sigma^2} -\frac{m_p^2}{N}\Big) \Psi_y \nonumber \\
&&+\frac{L^2 \lambda}{r^2 N} f^2 \Psi_p^2 \Psi_y  - \frac{L^2 q_p \gamma H f^2 \Psi_x }{r^2 N}.
\end{eqnarray}
The above equations admits the following four sets of scaling symmetries
\begin{eqnarray}
(1)& \phi \rightarrow \lambda^2 \phi~,~\Psi_x \rightarrow \lambda^2 \Psi_x~,~\Psi_y \rightarrow \lambda^2 \Psi_y~,~m_p\rightarrow \lambda m_p~,\nonumber\\&
N \rightarrow \lambda^2 N~,~L\rightarrow \lambda^{-1} L~,~b \rightarrow \lambda^{-1} b~;
\\(2)& \phi \rightarrow \lambda \phi~,~\Psi_x \rightarrow \lambda \Psi_x~,~\Psi_y \rightarrow \lambda \Psi_y~,~N \rightarrow \lambda^2 N~,\nonumber\\&
~M \rightarrow \lambda^3 M~,~r \rightarrow \lambda r~;
\\(3)& \phi \rightarrow \lambda \phi~,~q_p \rightarrow \lambda^{-1} q_p~,\\
&\Psi_x \rightarrow \lambda \Psi_x~,~\Psi_y \rightarrow \lambda \Psi_y~,~\Psi_p \rightarrow \lambda \Psi_p~. \label{scaling3}
\\(4)& \phi \rightarrow \lambda \phi~,~\sigma \rightarrow \lambda \sigma~; \label{scaling4}
\\(5)& \Psi_x \rightarrow \lambda \Psi_x~,~\Psi_y \rightarrow \lambda^{-1} \Psi_y~,~f \rightarrow \lambda^{-1} f~. \label{scaling5}
\end{eqnarray}
These scaling symmetries will be used to facilitate the numerical work. For instance, we use the first scaling symmetry to set $L=1$
and the second one to set $r_h=1$. After getting the numerical solutions, we use the two scaling symmetries again to
recover $L$ and $r_h$ to any value. The third scaling symmetry is used to scale $q_p$ to any value. Therefore in the rest of this paper, we set $L=r_h=q_p=1$ without lose of generality. The last two scaling symmetries are used to scale any solution to be asymptotically AdS, which means $\lim\limits_{r\to\infty} \sigma(r) \rightarrow1$ and $\lim\limits_{r\to\infty}f(r)\rightarrow 1$.

For the purpose of solving the equations of motion numerically, we need to specify the boundary conditions both
on the horizon $r=r_h=1$ and on the boundary $r=\infty$. 
Near the horizon the functions can be expanded as
\begin{eqnarray}
M(r) &=& \frac{r_h^3}{2}+M_{h1} (r-1) + ...~,\\
\sigma(r) &=& \sigma_{h0}+\sigma_{h1}(r-1)+...~,\\
f(r) &=& f_{h0} + f_{h1}(r-1)+...~,\\
\phi(r) &=& \phi_{h1}(r-1)+\phi_{h2}(r-1)^2+...~,\\
\Psi_x(r) &=& \Psi_{xh0}+\Psi_{xh1}(r-1)+...~,\\
\Psi_y(r) &=& \Psi_{yh0}+\Psi_{yh1}(r-1)+...~.\\
\Psi_p(r) &=& \Psi_{ph0}+\Psi_{ph1}(r-1)+...~.
\end{eqnarray}
One can check that only the coefficients \{$\sigma_{h0}$,$f_{h0}$,$\phi_{h1}$,\\$\Psi_{xh0}$,$\Psi_{yh0}$,$\Psi_{ph0}$\} are independent. We also need to expand the functions near the AdS boundary
\begin{eqnarray}
M(r) &=& M_{b0}+ \frac{M_{b1}}{r}+ ...~,\\
\sigma(r) &=& \sigma_{b0}+ \frac{\sigma_{b3}}{r^3}+ ...~,\\
f(r) &=& f_{b0}+ \frac{f_{b3}}{r^3}+ ...~,\\
\phi(r) &=& \mu- \frac{\rho}{r}+ ...~,\\
\Psi_x(r) &=& \frac{\Psi_{x-}}{r^{1-\triangle}}+\frac{\Psi_{x+}}{r^{\triangle}}+ ...~,\\
\Psi_y(r) &=& \frac{\Psi_{y-}}{r^{1-\triangle}}+\frac{\Psi_{y+}}{r^{\triangle}}+ ...~,\\
\Psi_p(r) &=& \frac{\Psi_{p-}}{r^{1-\triangle}}+\frac{\Psi_{p+}}{r^{\triangle}}+ ...~,
\end{eqnarray}
where
\begin{eqnarray}
\triangle &=& (1+\sqrt{1+4m_p^2 L^2})/2
\end{eqnarray}
is the operator dimension for both the p-wave and p+ip orders.

The AdS/CFT dictionary tells us that $\mu$ and $\rho$ are related to the chemical potential and charge density respectively. We choose the standard quantization, which means $\Psi_{x-}$, $\Psi_{y-}$, $\Psi_{p-}$ are related to the sources and $\Psi_{x+}$, $\Psi_{y+}$, $\Psi_{p+}$ are related to the expectation values of the dual vector operators. Since we focus on the solutions with no source term, we further set $\Psi_{x-}=\Psi_{y-}=\Psi_{p-}=0$ as three additional constrains.
With these three constraints together with $\sigma_{b0}=f_{b0}=1$, which keep the solutions to be asymptotical AdS, we obtain a set of one parameter solutions that mimic the p-wave or p+ip phase transition holographically.

Besides the normal(R-N) solution with $\Psi_x=\Psi_y=\Psi_p=0$ which is dual to the normal phase, this holographic model also permits the p-wave solution with $\Psi_x=\Psi_y=0,~\Psi_p\neq 0$ and the p+ip solution with $\Psi_x=\Psi_y\neq 0,~\Psi_p=0$. Both the p-wave and the p+ip solutions are dual to superfluid phases which exist below the respective critical temperatures and have lower grand potential than the normal solution.

In the p-wave solution, the condensate value can be extracted by the expectation value $\langle O\rangle=\Psi_{p+}$. While in the p+ip solution, the condensate value involve with two expectation values $\Psi_{x+}$ and $\Psi_{y+}$. In order to compare the condensate value of the p-wave and p+ip solutions, we define the condensate value of the p+ip solution as
\begin{eqnarray}
\langle O_{\text{pip}}\rangle &=& \sqrt{\Psi^2_{x+}+\Psi^2_{y+}},
\end{eqnarray}
which is in consistent with the fact that the p-wave and the p+ip solutions are equal stable and get the same condensate value in probe limit~\cite{Nie:2016pjt}.

In order to find out which one between the two condensed solutions is the most stable, we work in grand canonical ensemble and calculate their grand potential $\Omega$, which can be evaluated by the on-shell Euclidean action of the bulk system. The final formula can be expressed as
\begin{eqnarray}
\frac{2\kappa_g^2}{V_2} \Omega=&&- \int_{r_h}^\infty \Big( -\frac{2 b^2 L^2 \sigma H}{r^2} + \frac{4b^2L^2 q_p\gamma \sigma H \Psi_x\Psi_y}{r^2} \Big)dr
\nonumber \\
&& -\frac{2 M|_{r\rightarrow \infty}}{L^4}-\frac{b^2 L^2 H^2}{2 r_h}+\frac{6 r_h^3\sigma'''|_{r\rightarrow \infty}}{L^4}.
\end{eqnarray}

With the above formula, we solve the equations and compare the stability of different solutions at different values of temperature $T$, magnetic intensity $H$ and back-reaction parameters $b$. Finally the 3-D $b-H-T$ phase structure is captured. Before we introduce two typical $H-T$ phase diagrams with two different values of $b$ to show the key property of the $b-H-T$ phase structure, we analyze the influence of $b$ and $H$ on the p-wave and p+ip solutions to better understand the final phase structure.
\section{Review the influence of back-reaction $b$}
\label{sect:b}
\begin{figure}
\includegraphics[width=0.49\columnwidth] {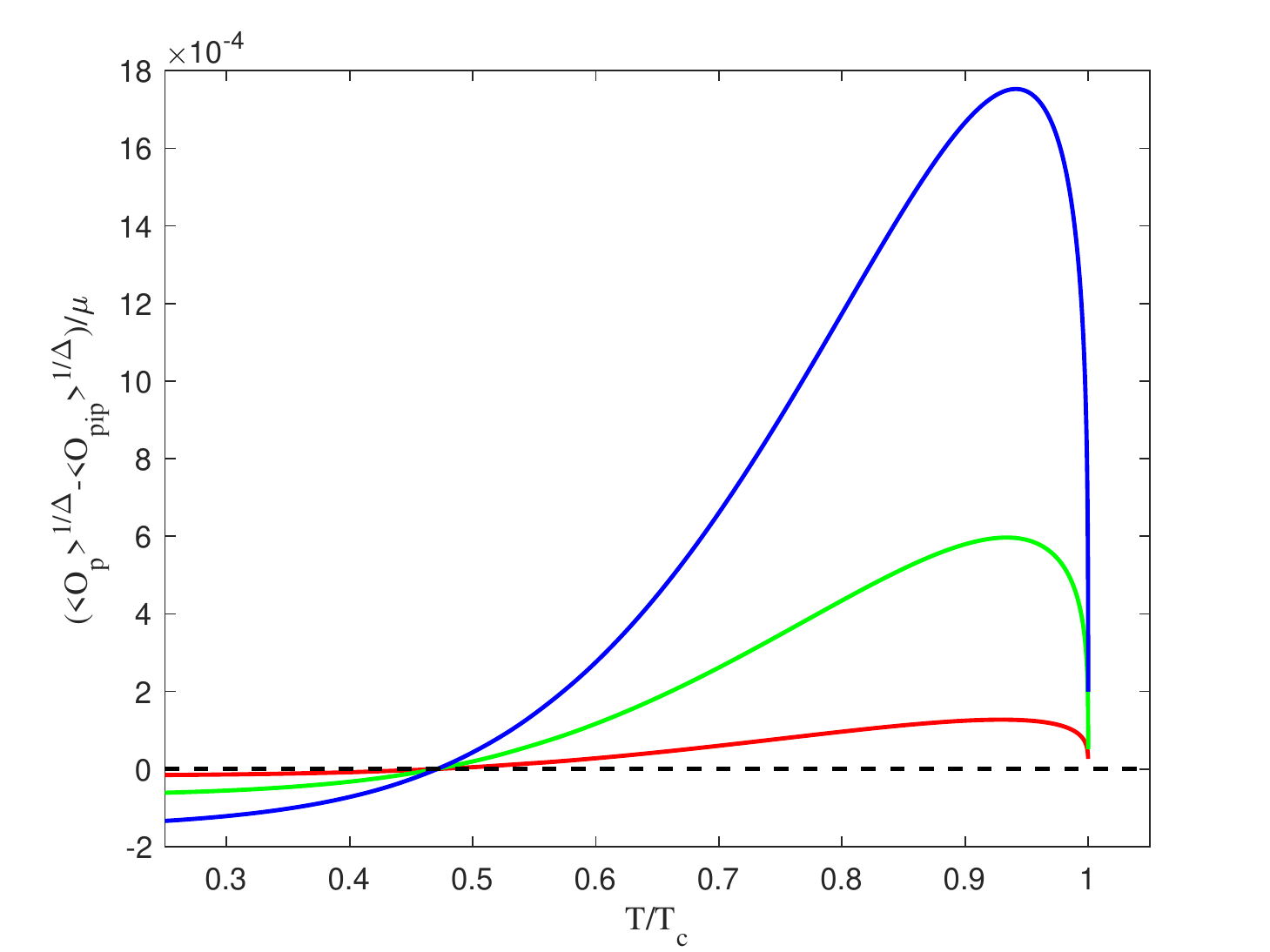}
\includegraphics[width=0.49\columnwidth] {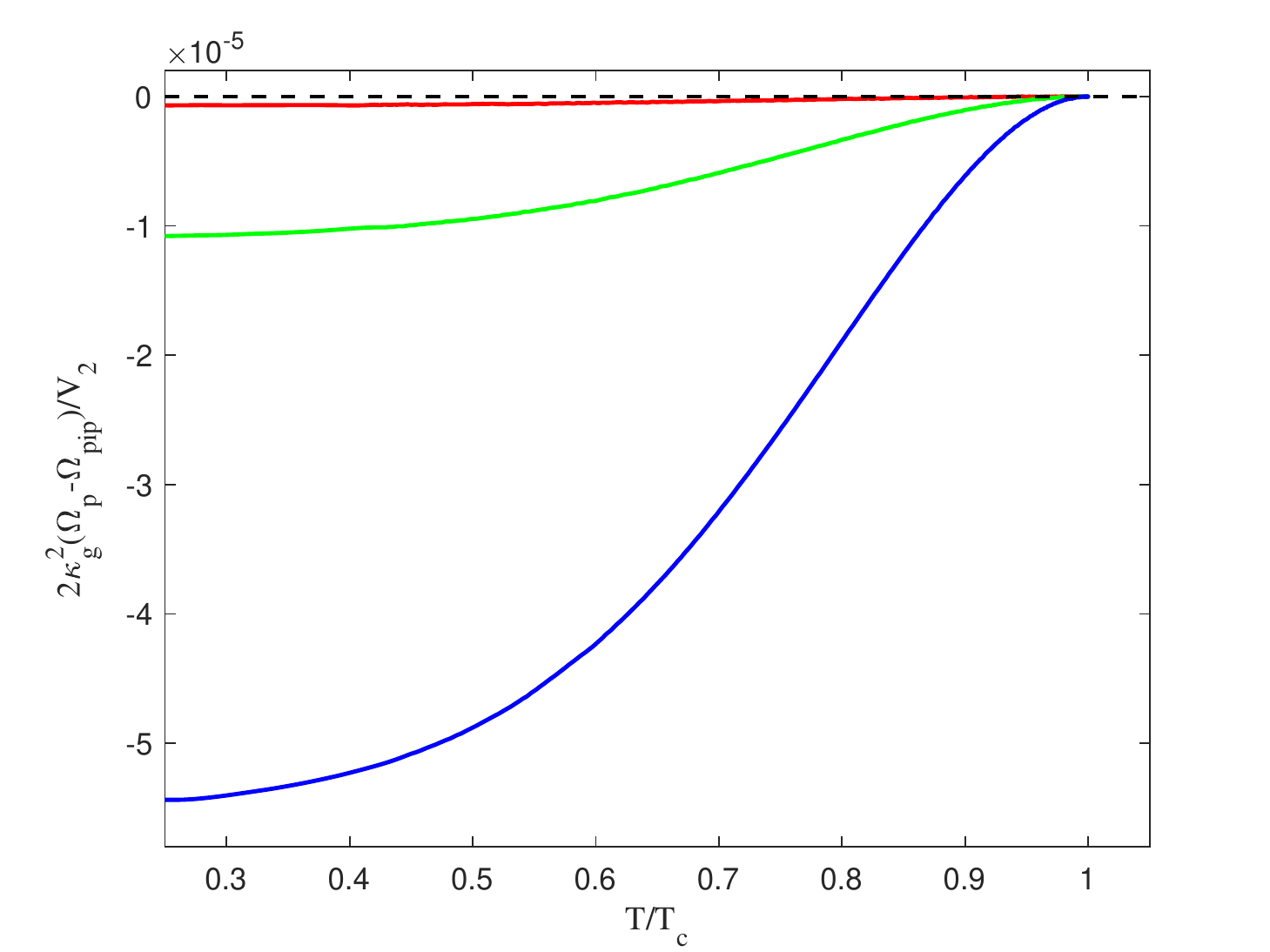}
\caption{\label{differentb}Difference of condensate(left plot) and grand potential density(right plot) between the p-wave and p+ip solutions with $b=0.1$(red line), $0.2$(green line), $0.3$(blue line).}
\end{figure}

The influence of back-reaction parameter $b$ on the difference between p-wave and p+ip solutions is already studied in Ref.~\cite{Nie:2016pjt}. In this section, we briefly review these results including the effects of back-reaction parameter $b$ on the difference of condensate as well as grand potential density between the two solutions, as represented in Figure.~\ref{differentb}.

The p-wave and p+ip solutions are the same stable and form degenerate states in probe limit. When back reaction is considered, the p-wave solution and p+ip solution still shall the same critical point where the infinitesimal condensates meet the conception of the probe limit. Therefore the condensates as well as grand potential of the two solutions can be conveniently compared with a horizontal axis $T/T_c$.

In the left plot of Figure.~\ref{differentb}, we plot the difference between the condensates of the p-wave and p+ip solutions $(\sqrt[\Delta]{\langle O_{\text{p}}\rangle}-\sqrt[\Delta]{\langle O_{\text{pip}}\rangle})$ versus temperature $T/T_{c}$ with $m^2_{p}=0$, $H=0$ and three values of back reaction parameter $b=0.1$(red),~$0.2$(green),~$0.3$(blue). We can see from this plot that the condensate value of the p-wave solution is different with that of p+ip solution. When the back reaction strength increase, the difference between the condensate values of p-wave and p+ip solutions become larger, which is consistent with the result in the probe limit that the condensate values of the p-wave and p+ip solutions are equal. Another interesting issue is that the condensate value of the p-wave solution is not always larger than that of the p+ip solution, and the three curves showing the difference of condensate values cross the horizontal axis at almost the same point, indicating coincidence or possible laws that need further investigation.

In the right plot of Figure.~\ref{differentb}, we plot the relative values of grand potential density $(\Omega_{p}-\Omega_{pip})/V_2$ versus temperature $T/T_{c}$ with $m^2_{p}=0$, $H=0$ and three values of back reaction parameter $b=0.1$(red),~$0.2$(green),~$0.3$(blue). As shown in this plot, the p-wave solution always get a lower value of grand potential, and the norm of the difference between the p-wave and p+ip solutions become larger when back reaction parameter increases or the temperature decreases.
\section{Influence of the magnetic field H}
\label{sect:H}
The influence of $H$ on the p-wave and p+ip solutions is quite interesting and crucial, because the p+ip solution is chiral and has to get nontrivial influence from coupled magnetic field.

As studied in Ref.~\cite{Nie:2016pjt}, the p+ip solution is isotropic in x-y plane, while the p-wave solution choose a special direction, which can be confirmed from the energy momentum tensor. The p+ip solution is a chiral state, in the sense that if we make a mirror transformation $y\rightarrow -y$, we get a ``p-ip'' solution. Both the p+ip and the ``p-ip'' solutions get the same condensate value as well as grand potential with zero magnetic intensity $H=0$. When coupling to a non zero magnetic field $H=M_{xy}$ is considered, the symmetry between the two chiral states are broken, therefore one of the two solutions become more stable while the other solution become less stable.

The value of $H$ can also be negtive, however, the discrete symmetry in the equations of motion $\Psi_y\rightarrow-\Psi_y, H\rightarrow-H$ tells us that the p+ip solution with $H>0$ is equivalent to the ``p-ip'' one with $H<0$ (in the sense they get the same condensate as well as grand potential), therefore we only consider the p+ip solution with both positive and negative values of $H$ is this paper.

\begin{figure}
\includegraphics[width=0.9\columnwidth] {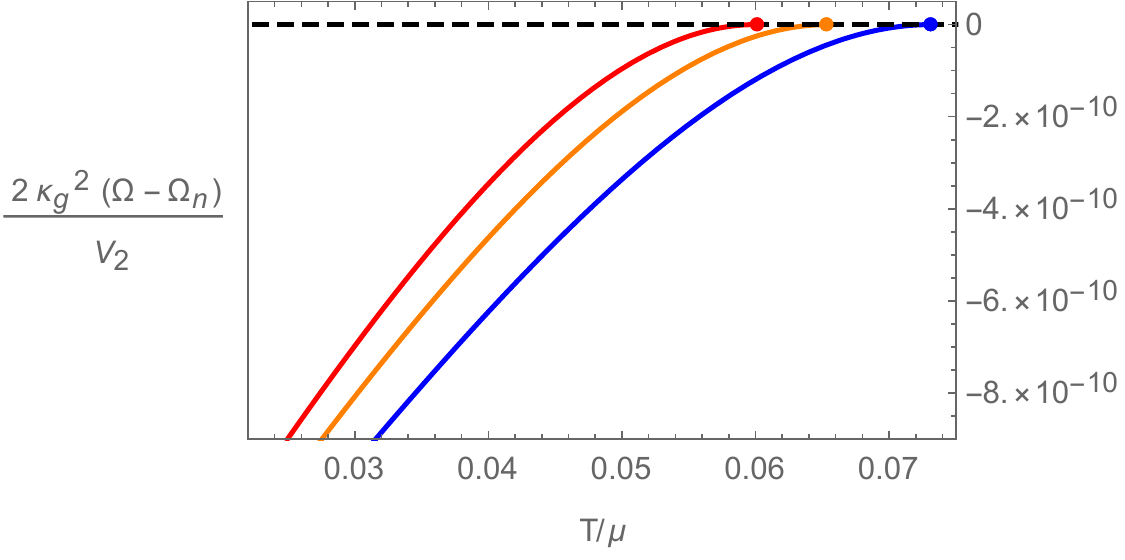}
\caption{\label{FreeE-testH}The relative value of grand potential density with respect to the normal phase in the case of $b=0.0001$ and $H=-0.5$(red line), $0$(orange line), $0.5$(blue line), with the three colored points mark the critical points.}
\end{figure}

In Figure \ref{FreeE-testH}, we fix the back-reaction parameter $b=0.0001$, set $m^2_{p}=0$ and draw the relative value of grand potential density between the p+ip solution and the normal solution $(\Omega_{pip}-\Omega_{normal})/V_2$ versus temperature $T/\mu$ with magnetic field intensity $H=-0.5$, $0$, and $0.5$, respectively. We can see that when the magnetic field intensity $H$ increases from -0.5 to 0.5, the critical point shifts rightwards, from red curve to blue curve. The grand potential density curve is also entirely ``parallel'' shifted. The p+ip solution get a higher critical temperature as well as lower grand potential density when $H$ is larger, therefore the p+ip solution becomes more stable with a larger value of $H$.

The value of $b$ is very small and therefore the results of Figure \ref{FreeE-testH} also indicate the results in probe limit. For general values of back reaction parameter $b$, we have checked that the results of the influence of $H$ are qualitative the same.

In this section, we do not show the influence of $H$ on the p-wave solution, for that the p-wave order is not directly coupled to the magnetic field and only get very limited influence that can be ignored while considering the qualitative feature of thee final phase diagram.
\section{Phase diagram}\label{sect:PD}
From the results in the above two sections, we can see that the influence of back reaction parameter $b$ is enlarging the difference between the grand potential of the two solutions sharing the same critical point, while the influence of $H$ is shifting the entire grand potential density curve of p+ip solution with the critical point also shifted. If we combine the two effects together, we get the interesting phase structure of this holographic model. We introduce two typical $H-T$ phase diagrams with two different values of $b$ as the main results to show the key property of the $b-H-T$ phase structure.

\begin{figure}
\includegraphics[width=0.49\columnwidth] {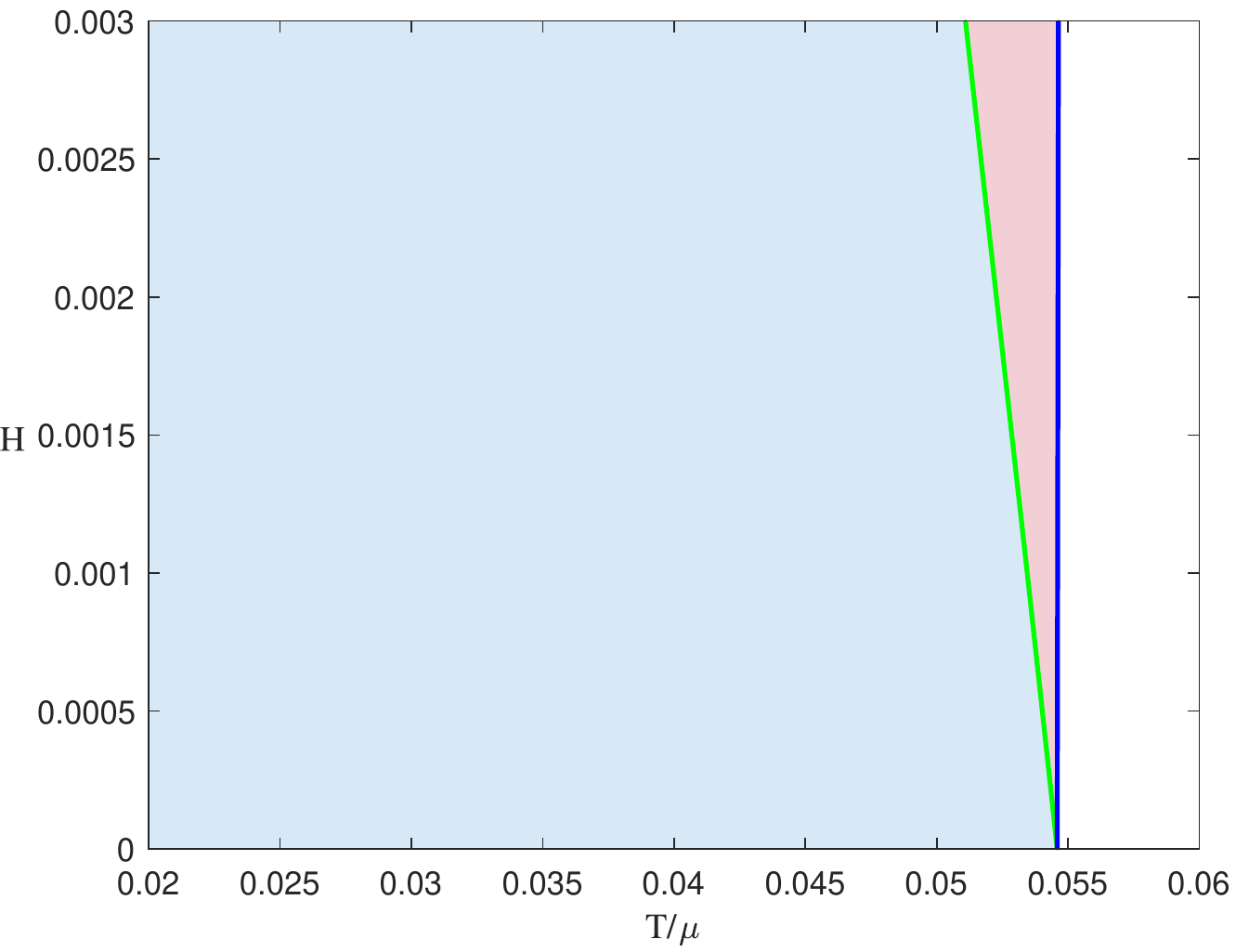}
\includegraphics[width=0.49\columnwidth] {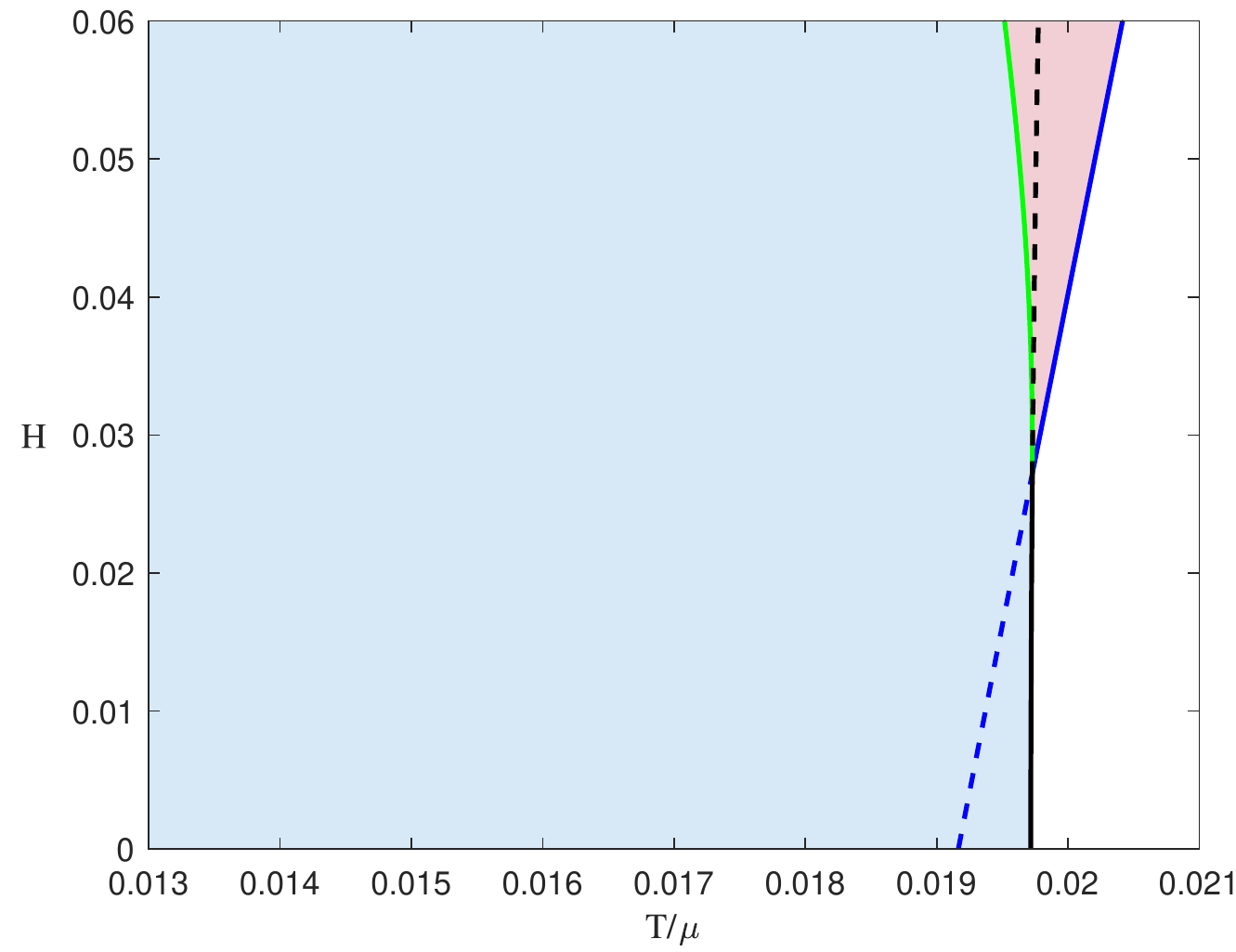}
\caption{\label{HTPDb03}The $H-T$ phase diagrams with $b=0.3$(left plot) and $b=0.68$(right plot). The region colored light blue, light red and white denote the p-wave phase, the p+ip phase and the normal phase, respectively. The solid blue line denotes the critical points of second order phase transition between the normal phase and the p+ip phase, the solid green line denotes the first order phase transition points between the p+ip phase and the p-wave phase. In the right plot, the solid black line denotes the first order phase transition points between the normal phase and the p-wave phase, and the dashed part of the blue line denotes the ``critical point'' of the p+ip solution if the p-wave solution is not included. The tricritical Point in the right plot is $(T_{t}=0.01973,H_{t}=0.02720)$.}
\end{figure}

In Figure.~\ref{HTPDb03}, we draw the two $H-T$ phase diagrams with $b=0.3$ (in the left plot) and $b=0.68$ (in the right plot), respectively. In these two phase diagrams, we use light blue, light red and white to mark the regions dominated by the p-wave phase, the p+ip phase and the normal phase, respectively. The solid blue line between the white and light red region denotes the critical points of second order phase transition between the normal phase and the p+ip phase. The dashed part of the blue line in the right plot also denotes the ``critical point'' of the p+ip solution, however, at these points, a p-wave solution with lower grand potential already exist and these points on the dashed blue line are not real phase transition points. The solid green line between the light blue and light red region denotes the first order phase transition points between the p+ip phase and the p-wave phase. The solid black line in the right plot denotes the first order phase transition points between the normal phase and the p-wave phase. There is also a dashed segment for the black line, which denotes the intersection points of grand potential curves of normal and p-wave solutions. Because the same reason as that of the dashed blue line, The points on the dashed part of the black line are not real phase transition points.

We should notice that near the solid green line, which denotes the intersection points of grand potential curves of p-wave and p+ip solutions, a coexistent solution with both the two orders duel to $\rho_{1\mu}$ and $\rho_{2\mu}$ non-zero might be the most stable. However, we can always increase the value of $\lambda$, which is the coefficient of the interacting term, to make the coexistent solution unstable~\cite{Zhang:2021vwp}. Therefore, the first order phase transitions between the p-wave and p+ip solutions marked by the solid green line is true with a large enough value of $\lambda$.

There is a sharp slit region in both the two phase diagrams. The difference is that the slit region in the left plot with $b=0.3$ starts from the critical point of the p-wave phase with zero magnetic intensity $H=0$, while the slit region in the right plot with $b=0.68$ starts from a tricritical point with finite magnetic intensity $H=H_t=0.02720$.

To understand this difference, we combine the effects of back-reaction parameter $b$ and the magnetic field $H$ on the grand potential density curves of p-wave and p+ip solutions. In the case with $b=0.3$, both the p-wave and p+ip phase transitions are second order and share the same critical point at zero magnetic intensity $H=0$. Moreover, the two solutions are equally stable at the critical point, and the grand potential density of the p-wave solution becomes smaller than that of the p+ip solution with finite condensate. With the magnetic field turned on, the critical point of p+ip solution moving rightwards along the increasing of $H$, while the critical point of the p-wave solution is only slightly affected. Therefore a stable region gradually emerge between the normal phase and p-wave phase and form the slit region in phase diagram. The slit region begin from the critical point with $H=0$, because an infinitesimal magnetic field cause an infinitesimal increase of the critical temperature of the p+ip solution, which becomes infinitesimally larger than the critical temperature of the p-wave solution.

The main difference of the case with $b=0.68$ is that the slit region begin from the tri-critical point with finite value of $H$. This is because that when $b>b_1=0.62$~\cite{Ammon:2009xh,Nie:2016pjt}, the p-wave phase transition with $H=0$ becomes first order. At this time, the phase transition temperature of the p-wave solution is always larger than that of the p+ip solution~\cite{Nie:2016pjt}. Therefore, when magnetic field is turned on, we need a finite magnetic intensity $H_t$ to shift the critical temperature of the p+ip solution and make it catching up with the critical temperature of the p-wave solution. Only if $H>H_t$, the critical temperature of the p+ip solution is large than that of the p-wave solution, and the slit region for stable p+ip solution emerge from the first order phase transition point of the p-wave phase at $H=H_t$. If the back-reaction parameter becomes larger than $b_2=0.69$, the p+ip phase transition also becomes first order~\cite{Nie:2016pjt}, in that case, most of the qualitative features of the $H-T$ phase diagram are still the same with the right plot in Figure.~\ref{HTPDb03}, the only difference is that the blue line denoting the p+ip phase transition becomes first order. The qualitative features of the two typical phase diagrams are quite stable. All the H-T phase diagrams with $b<b_1=0.62$ is similar to the left plot while all the $H-T$ phase diagrams with $b>b_1=0.62$ is similar to the right one.

The above results indicates a possible universality that \textbf{the zero (finite) value of some applied field of the start point of the slit region indicates the phase transition with vanishing applied field is second order (first order). This universality should be satisfied in a system that has a degenerate critical point (spinodal point) at zero applied field which break this degeneracy with finite value.}

The slit region feature of the phase diagram reminiscent the $H-T$ phase diagram of He-3, where a slit region of A1 phase always exist between the normal phase and the A2(or B) phase at different pressure. This is in consistent with that the phase transition from the normal phase to the A-phase(or B-phase) is second order in Helium-3, and the universality we claimed further indicates degenerate of the critical point of A(or B) phase at zero magnetic intensity. In addition, we can see that even two slit regions occur in the case with low pressure where B phase dominate the whole superfluid region with zero magnetic field. According to the universality we claimed, the two slit regions indicates higher degenerate structure of order parameters at the superfluid phase transition critical point at zero magnetic field.
\section{\bf Conclusions and discussions}\label{sect:conclusion}
In this work, we study the competition between the p-wave solution and the p+ip solution in a holographic model with two charged vector fields in 3+1 dimensional bulk. We reviewed the influence of back-reaction parameter and studied the influence of magnetic intensity $H$ on the phase structure. Combining the two influences, we get two typical $H-T$ phase diagrams with two values of back-reaction parameter $b=0.3$ and $b=0.68$. We compared the difference of the slit region in the two phase diagrams and find it is related to the order of the phase transition at zero magnetic field. We claim possible universality of this result and find it consistent with the Helium-3 $H-T$ phase diagrams.

Our results show interesting relation between slit regions in phase diagrams and degeneracy of phase transition critical points at zero applied field. It also indicates a clue of the order of phase transitions. We expect further investigations on this topic in future.
\section*{Acknowledgements}
ZYN would like to thank Professor Jan Zaanen and Professor Grisha Volovik for useful discussions and suggestions. This work was supported in part by the National Natural Science Foundation of China under Grant Nos. 11965013, 11565017, and 11675140, ZYN is partially supported by Yunnan Ten Thousand Talents Plan Young and Elite Talents Project.

\begin{thebibliography}{99}
\baselineskip 12pt
\bibitem{Maldacena:1997re}
  J.~M.~Maldacena,
  \emph{The large N limit of superconformal field theories and supergravity},
  Adv.\ Theor.\ Math.\ Phys.\  {\bf 2}, 231 (1998)
  [Int.\ J.\ Theor.\ Phys.\  {\bf 38}, 1113 (1999)]
  [arXiv:hep-th/9711200].
\bibitem{Gubser:1998bc}
  S.~S.~Gubser, I.~R.~Klebanov and A.~M.~Polyakov,
  \emph{Gauge theory correlators from non-critical string theory
  },
  Phys.\ Lett.\  B {\bf 428}, 105 (1998)
  [arXiv:hep-th/9802109].
\bibitem{Witten:1998qj}
  E.~Witten,
  \emph{Anti-de Sitter space and holography},
  Adv.\ Theor.\ Math.\ Phys.\  {\bf 2}, 253 (1998)
  [arXiv:hep-th/9802150].
\bibitem{Gubser:2008px}
  S.~S.~Gubser,
  \emph{Breaking an Abelian gauge symmetry near a black hole horizon},
  Phys.\ Rev.\  D {\bf 78}, 065034 (2008)
  [arXiv:0801.2977 [hep-th]].
\bibitem{Hartnoll:2008vx}
  S.~A.~Hartnoll, C.~P.~Herzog and G.~T.~Horowitz,
  \emph{Building a Holographic Superconductor},
  Phys.\ Rev.\ Lett.\  {\bf 101}, 031601 (2008)
  [arXiv:0803.3295 [hep-th]].
\bibitem{Herzog:2008he}
  C.~P.~Herzog, P.~K.~Kovtun and D.~T.~Son,
  \emph{Holographic model of superfluidity},
  Phys.\ Rev.\ D {\bf 79}, 066002 (2009)
  [arXiv:0809.4870 [hep-th]].
\bibitem{Zaanen:2015oix}
  J.~Zaanen, Y.~W.~Sun, Y.~Liu and K.~Schalm,
  \emph{Holographic Duality in Condensed Matter Physics},
  Cambridge Univ. Press (11-2015)
\bibitem{Gubser:2008wv}
  S.~S.~Gubser and S.~S.~Pufu,
  \emph{The gravity dual of a p-wave superconductor},
  JHEP {\bf 0811} (2008) 033
  [arXiv:0805.2960 [hep-th]].
\bibitem{Cai:2013aca}
  R.~G.~Cai, L.~Li and L.~F.~Li,
  \emph{A Holographic P-wave Superconductor Model},
  JHEP {\bf 1401}, 032 (2014)
  [arXiv:1309.4877 [hep-th]].







\bibitem{Cai:2013pda}
  R.~G.~Cai, S.~He, L.~Li and L.~F.~Li,
  \emph{A Holographic Study on Vector Condensate Induced by a Magnetic Field},
  JHEP {\bf 1312}, 036 (2013)
  [arXiv:1309.2098 [hep-th]].
\bibitem{Li:2013rhw}
  L.~F.~Li, R.~G.~Cai, L.~Li and C.~Shen,
  \emph{Entanglement entropy in a holographic p-wave superconductor model},
  Nucl.\ Phys.\ B {\bf 894}, 15 (2015)
  [arXiv:1310.6239 [hep-th]].
\bibitem{Wu:2014bba}
  Y.~B.~Wu, J.~W.~Lu, W.~X.~Zhang, C.~Y.~Zhang, J.~B.~Lu and F.~Yu,
  \emph{Holographic $p$-wave superfluid},
  Phys.\ Rev.\ D {\bf 90}, no. 12, 126006 (2014)
  [arXiv:1410.5243 [hep-th]].
\bibitem{Rogatko:2015nta}
M.~Rogatko and K.~I.~Wysokinski,
  \emph{P-wave holographic superconductor/insulator phase transitions affected by dark matter sector},
JHEP \textbf{03}, 215 (2016)
[arXiv:1508.02869 [hep-th]].
\bibitem{Pan:2015lit}
  Q.~Pan and S.~J.~Zhang,
  \emph{Revisiting holographic superconductors with hyperscaling violation},
  Eur.\ Phys.\ J.\ C {\bf 76}, no. 3, 126 (2016)
  [arXiv:1510.09199 [hep-th]].
\bibitem{Lai:2016yma}
  C.~Lai, Q.~Pan, J.~Jing and Y.~Wang,
  \emph{Analytical study on holographic superfluid in AdS soliton background},
  Phys.\ Lett.\ B {\bf 757}, 65 (2016)
  [arXiv:1601.00134 [hep-th]].
\bibitem{Nie:2016pjt}
  Z.~Y.~Nie, Q.~Pan, H.~B.~Zeng and H.~Zeng,
 \emph{Split degenerate states and stable p $+i$ p phases from holography},
  Eur.\ Phys.\ J.\ C {\bf 77}, no. 2, 69 (2017)
  [arXiv:1611.07278 [hep-th]].
\bibitem{Lu:2018tdo}
  J.~W.~Lu, Y.~B.~Wu, B.~P.~Dong and H.~Liao,
  \emph{Holographic p-wave superconductor in Lifshitz gravity with $RF^2$ correction},
  Phys.\ Lett.\ B {\bf 785}, 517 (2018).
\bibitem{Huang:2019yov}
  Y.~Huang, Q.~Pan, W.~L.~Qian, J.~Jing and S.~Wang,
  \emph{Holographic p-wave superfluid with Weyl corrections},
  Sci.\ China Phys.\ Mech.\ Astron.\  {\bf 63}, no. 3, 230411 (2020).
\bibitem{Lu:2020phn}
  J.~W.~Lu, Y.~B.~Wu, B.~P.~Dong and Y.~Zhang,
  \emph{Holographic p-wave superconductor with $C^2F^2$ correction},
  Eur.\ Phys.\ J.\ C {\bf 80}, no. 2, 114 (2020).
\bibitem{Lv:2020ecm}
  Y.~Lv, X.~Qiao, M.~Wang, Q.~Pan, W.~L.~Qian and J.~Jing,
  \emph{Holographic p-wave superfluid in the AdS soliton background with RF 2 corrections},
  Phys.\ Lett.\ B {\bf 802}, 135216 (2020)
  [arXiv:2001.08364 [hep-th]].
\bibitem{Nie:2020lop}
Z.~Y.~Nie, Y.~P.~Hu and H.~Zeng,
Eur. Phys. J. C \textbf{80}, no.11, 1015 (2020)
doi:10.1140/epjc/s10052-020-08594-4
[arXiv:2003.12989 [hep-th]].
\bibitem{Qiao:2020hkx}
X.~Qiao, L.~OuYang, D.~Wang, Q.~Pan and J.~Jing,
JHEP \textbf{12}, 192 (2020)
doi:10.1007/JHEP12(2020)192
[arXiv:2005.01007 [hep-th]].
\bibitem{Lu:2021tln}
J.~W.~Lu, Y.~B.~Wu, H.~F.~Li, B.~P.~Dong, Y.~Zheng and H.~Liao,
Phys. Lett. B \textbf{819}, 136448 (2021)
doi:10.1016/j.physletb.2021.136448




















\bibitem{Nie:2013sda}
Z.~Y.~Nie, R.~G.~Cai, X.~Gao and H.~Zeng,
JHEP \textbf{11}, 087 (2013)
doi:10.1007/JHEP11(2013)087
[arXiv:1309.2204 [hep-th]].
\bibitem{Amado:2013lia}
I.~Amado, D.~Arean, A.~Jimenez-Alba, L.~Melgar and I.~Salazar Landea,
Phys. Rev. D \textbf{89}, no.2, 026009 (2014)
doi:10.1103/PhysRevD.89.026009
[arXiv:1309.5086 [hep-th]].
\bibitem{Momeni:2013bca}
D.~Momeni, M.~Raza and R.~Myrzakulov,
Int. J. Geom. Meth. Mod. Phys. \textbf{12}, no.04, 1550048 (2015)
doi:10.1142/S0219887815500486
[arXiv:1310.1735 [hep-th]].
\bibitem{Nie:2014qma}
Z.~Y.~Nie, R.~G.~Cai, X.~Gao, L.~Li and H.~Zeng,
Eur. Phys. J. C \textbf{75}, 559 (2015)
doi:10.1140/epjc/s10052-015-3773-2
[arXiv:1501.00004 [hep-th]].
\bibitem{Liu:2015zca}
S.~Liu and Y.~Q.~Wang,
Eur. Phys. J. C \textbf{75}, no.10, 493 (2015)
doi:10.1140/epjc/s10052-015-3692-2
[arXiv:1504.06918 [hep-th]].
\bibitem{Nie:2015zia}
Z.~Y.~Nie and H.~Zeng,
JHEP \textbf{10}, 047 (2015)
doi:10.1007/JHEP10(2015)047
[arXiv:1505.02289 [hep-th]].
\bibitem{Wang:2016jov}
Y.~Q.~Wang and S.~Liu,
JHEP \textbf{11}, 127 (2016)
doi:10.1007/JHEP11(2016)127
[arXiv:1608.06364 [hep-th]].
\bibitem{Li:2017wvw}
R.~Li, T.~Zi and H.~Zhang,
Phys. Lett. B \textbf{766}, 238-244 (2017)
doi:10.1016/j.physletb.2017.01.018
\bibitem{Momeni:2019kfl}
D.~Momeni, N.~Majd, M.~Mohammadzaheri, P.~Channuie and M.~Al Ajmi,
Results Phys. \textbf{14}, 102449 (2019)
doi:10.1016/j.rinp.2019.102449
[arXiv:1908.07994 [hep-th]].
\bibitem{Xia:2021pap}
C.~Y.~Xia, Z.~Y.~Nie, H.~B.~Zeng and Y.~Zhang,
Eur. Phys. J. C \textbf{81}, no.10, 882 (2021)
doi:10.1140/epjc/s10052-021-09684-7
[arXiv:2102.01083 [hep-th]].
\bibitem{Zhang:2021vwp}
X.~K.~Zhang, C.~Y.~Xia, Z.~Y.~Nie and H.~Zeng,
[arXiv:2105.14294 [hep-th]].
\bibitem{Volhardt-Wolfle-1990} D.~Vollhardt and P.~Wolfle,
 \emph{The superfluid phases of helium 3},
 Philadelphia, PA (USA); Taylor and Francis Inc. 1990.
\bibitem{Ammon:2009xh}
  M.~Ammon, J.~Erdmenger, V.~Grass, P.~Kerner and A.~O'Bannon,
  \emph{On Holographic p-wave Superfluids with Back-reaction},
  Phys.\ Lett.\ B {\bf 686}, 192 (2010)
  [arXiv:0912.3515 [hep-th]].

\end{thebibliography}

\end{document}